\documentclass[10pt, final]{scrartcl}
\pdfoutput=1

\usepackage[utf8]{inputenc}
\usepackage[T1]{fontenc}

\usepackage{microtype}
\usepackage{tabularx}
\usepackage{booktabs}
\usepackage{color}
\usepackage[english]{babel}

\usepackage{slashed}
\usepackage{caption}
\usepackage{subcaption}
\usepackage{amsmath, amssymb}
\usepackage{amsthm}
\usepackage{xspace}
\usepackage{graphicx}

\newcommand{\rnum}[1]{\uppercase\expandafter{\romannumeral #1\relax}}

\newcommand{\fer}[1]{\psi_{#1}}
\newcommand{\afer}[1]{\overline{\psi}_{#1}}
\newcommand{\gau}[1]{\lambda_{#1}}

\newcommand{\yuk}[2]{\ensuremath{\smash[t]{\Upsilon_{#1}^{\protect\phantom{#1}#2}}}\xspace}
\newcommand{\yuks}[2]{\ensuremath{\Upsilon_{#1}^{\phantom{#1}#2}}{}^*\xspace}

\newcommand{\yukw}[2]{\ensuremath{\widetilde{\Upsilon}_{#1}^{\protect\phantom{#1}#2}}\xspace}

\def\sfer{\widetilde{\psi}}
\def\asfer{\overline{\widetilde{\psi}}}

\newcommand{\BBBB}[1]{\ensuremath{\mathcal{B}_{\textrm{maj}}}\xspace} 
\newcommand{\B}[1]{\ensuremath{\mathcal{B}_{#1}}\xspace}

\newcommand{\Bc}[2]{\ensuremath{\mathcal{B}_{#1}^{#2}}\xspace}

\def\bas#1\eas{\begin{align*}#1\end{align*}}
\def\ba#1\ea{\begin{align}#1\end{align}}
\def\bi#1\ei{\begin{itemize}#1\end{itemize}}
\def\be#1\ee{\begin{enumerate}#1\end{enumerate}}
\def\nn{\nonumber}

\def\maj{\Upsilon_{\mathrm{m}}}

\def\A{\mathcal{A}}

\def\H{\mathcal{H}}

\def\n{\mathcal{N}}

\def\com{\mathbb{C}}

\DeclareMathOperator{\tr}{tr}
\DeclareMathOperator{\id}{id}

\DeclareMathOperator{\ad}{ad}

\newcommand{\rep}[2]{\ensuremath{\mathbf{N}_{#1} \otimes \mathbf{N}_{#2}^{o}}\xspace}
\newcommand{\repl}[2]{\ensuremath{\mathbf{#1} \otimes \mathbf{#2}^o}\xspace}

\def\dirac{\ensuremath{\slashed{\partial}_M}\xspace}

\newcommand{\w}[1]{\ensuremath{\omega_{#1}}}

\def\sgnc{\epsilon}

\theoremstyle{plain}
\newtheorem{theorem}{Theorem}

\usepackage{import}

\newtoks\svgpath
\svgpath={gfx/}

\graphicspath{{gfx/}}

\newcommand{\includesvg}[1]{
\subimport{\the\svgpath}{#1.pdf_tex}
}

\newcommand{\fig}[6]{\begin{figure}[#1] \centering \def\svgwidth{#2} \subimport{\the\svgpath}{#3.pdf_tex}\captionsetup{width=#4}\caption{#5}\label{fig:#6} \end{figure}}

\usepackage{a4wide}

\usepackage{parskip}
\def\sfer{\ensuremath{\phi}\xspace}
\def\asfer{\ensuremath{\bar\phi}\xspace}

\usepackage[]{showkeys}

\usepackage[affil-it]{authblk}
\author[a]{Wim Beenakker%
  \thanks{Electronic address: \texttt{W.Beenakker@science.ru.nl}}}
\author[a,b]{Thijs van den Broek%
  \thanks{Electronic address: \texttt{T.vandenBroek@science.ru.nl}}}
\author[a]{Walter D.~van Suijlekom%
  \thanks{Electronic address: \texttt{waltervs@math.ru.nl} (corresponding author)}}
\affil[a]{Radboud University Nijmegen, Institute for Mathematics, Astrophysics and Particle Physics,
Faculty of Science, PO Box 9010, 6500 GL, Nijmegen, The Netherlands}
\affil[b]{Nikhef, Science Park Amsterdam 105, 1098 XG Amsterdam}

\title{Supersymmetry and noncommutative geometry}
\subtitle{Part \rnum{2}: Supersymmetry breaking}
\date{}

\begin{document}

\maketitle

\begin{abstract}
We describe how a soft supersymmetry breaking Lagrangian arises naturally in the context of almost-commutative geometries that fall within the classification of those having a supersymmetric particle content as well as a supersymmetric spectral action. All contributions to such a Lagrangian are seen to either be generated automatically after introducing gaugino masses to the theory or coming from the second Seeley-DeWitt coefficient that is already part of the spectral action. In noncommutative geometry, a supersymmetric particle content and the appearance of a soft breaking Lagrangian thus appear to be intimately connected to each other.
\end{abstract}

\section{Introduction}

Already shortly after the advent of supersymmetry (e.g.~\cite{WestBook}) it was realized \cite{WZ74-2} that if it is a real symmetry of nature, then the superpartners 
should be of equal mass. This, however, is very much not the case. If it were, we should have seen all the sfermions and gauginos that feature in the Minimal Supersymmetric Standard Model (MSSM, e.g.~\cite{DGR04}) in particle accelerators by now. In the context of the MSSM we need \cite{GH86} a supersymmetry breaking Higgs potential to get electroweak symmetry breaking and give mass to the SM particles. Somehow there should be a mechanism at play that \emph{breaks} supersymmetry. Over the years many mechanisms have been suggested that break supersymmetry and explain why the masses of superpartners should be different at low scales. Ideally this should be mediated by a \emph{spontaneous} symmetry breaking mechanism, such as $D$-term \cite{R75} or $F$-term \cite{FI74} supersymmetry breaking. But phenomenologically such schemes are disfavoured, for they require that `in each family at least one slepton/squark is lighter than the corresponding fermion' \cite[\S 9.1]{DGR04}. Alternatively, supersymmetry can be broken \emph{explicitly} by means of a supersymmetry breaking Lagrangian. In order for the solution to the hierarchy problem that supersymmetry provides to remain useful, the terms in this supersymmetry breaking Lagrangian should be \emph{soft} \cite{GD81}. This means that such terms have couplings of positive mass dimension, not yield quadratically divergent loop corrections that would spoil the solution to the hierarchy problem (the enormous sensitivity of the Higgs boson mass to perturbative corrections) that supersymmetry provides. 

In \cite{BS13I} we provided a classification of potentially supersymmetric models within the framework of noncommutative geometry \cite{C94,CCM07}. The question on how to \emph{break} supersymmetry is naturally the next one to ask. In this paper, that can be considered as a follow-up on \cite{BS13I}, we will propose an answer to that question. The paper is organised as follows. First we will provide a short recapitulation of the aforementioned classification in Section \ref{sec:recap}. Next, we will shortly review \emph{soft supersymmetry breaking} in Section \ref{sec:Lsoft}. For each of the possible soft supersymmetry breaking terms we will ask ourselves in Section \ref{sec:terms} if and how it can also arise in noncommutative geometry. We find that all types of contributions that typically enter a soft supersymmetry breaking Lagrangian are also generated in this context. All terms are seen to be either generated by extra contributions to the action that arise after introducing gaugino masses or they come from the second Seeley-DeWitt coefficient that is already part of the spectral action. We show that the soft supersymmetry breaking terms that we get include all terms that can be generated via gaugino masses. 

\section{Supersymmetry in noncommutative geometry}\label{sec:recap}

The context in which the classification of potentially supersymmetric theories was found, was a particular class of noncommutative geometries; the \emph{almost-commutative geometries} (\cite{S00}, see \cite{DS11} for an introduction),
\begin{align*}
	(C^{\infty}(M, \A_F), L^2(M, S \otimes \H_F), i\gamma^\mu \nabla^S_\mu + \gamma_M \otimes D_F, J_M \otimes J_F; \gamma_M \otimes \gamma_F).
\end{align*}
It is the tensor product of a (real, even) \emph{canonical spectral triple} \cite[Ch 6.1]{C94} with a (real, even) \emph{finite spectral triple} \cite{KR97}. With the first we mean the data
\begin{align*}
	(C^{\infty}(M), L^2(M, S), \dirac = i\gamma^\mu \nabla^S_\mu; J_M, \gamma_M),
\end{align*}
where $(M, g)$ is a compact Riemannian spin manifold, $L^2(M, S)$ denotes the square integrable sections of the corresponding spinor bundle and \dirac is the \emph{Dirac operator} that is derived from the Levi-Civita connection on $M$, $\gamma_M$ is the chirality operator (only for even-dimensional $M$) and $J_M$ denotes charge conjugation. Real, even finite spectral triples are all of the form 
\begin{align*}
	(\A_F, \H_F, D_F; J_F, \gamma_F)   
\end{align*}
where $\A_F$ is a (finite) direct sum of matrix algebras over $\mathbb{R}$, $\mathbb{C}$ or $\mathbb{H}$, $\H_F$ is a (finite dimensional) $\A_F$-bimodule, whose right module structure is implemented by a \emph{real structure} $J_F$ (i.e.~$ \xi a := J_Fa^*J^*_F\xi$, $\xi \in \H_F$, $a \in \A_F$), $\gamma_F$ is a grading (i.e.~$\gamma^2_F = 1, \gamma_F = \gamma^*_F$) and $D_F$ is a Hermitian matrix on $\H_F$. \\

There are several extra demands on the elements of spectral triples, a couple of them we will list here. First of all, the Dirac operator must anticommute with the grading:
\begin{align}
	D\gamma = - \gamma D \label{eq:Dg-anti-comm}.
\end{align}
Secondly, if we can define a real structure $J$ (such as $J_M$ or $J_F$ above), the Dirac operator must be compatible with it via the \emph{order one condition}:
\begin{align}
[[D, a], Jb^*J^*] &= 0\ \forall\ a, b \in \A\label{eq:order_one}.
\end{align}
Finally, given $J$, there are the following relations:
\begin{align}
	J^2 &= \pm 1, & DJ &= \pm JD, &J\gamma &= \pm \gamma J. \label{eq:JD-anti-comm}
\end{align}
The three signs above give rise to the notion of KO-dimension which is defined modulo $8$. The signs for the even KO-dimensions are given in Table \ref{tb:ko-dimensions}.  The KO-dimension of a canonical spectral triple automatically equals the metric dimension of the manifold $M$ that it is defined on. The KO-dimension of the tensor product of two spectral triples is equal to the sum of their respective KO-dimensions \cite{Dabrowski2010}.\\
\begin{table}[h!]
\begin{tabularx}{\textwidth}{X cccc X}
			\toprule
  & KO-dimension & $J^2 = \epsilon$ & $JD = \epsilon' DJ$ & $J\gamma = \epsilon''\gamma J$ & 
\\
\midrule
        & 0 & + & + & +&\\ 
        & 2 & $-$ & + & $-$ &\\
        & 4 & $-$ & + & + &\\
        & 6 & + & + & $-$&\\
\bottomrule
\end{tabularx}
\caption{The possible values for the signs \cite[Ch 9.5]{GVF00} in \eqref{eq:JD-anti-comm} for the even KO-dimensions.}
\label{tb:ko-dimensions}
\end{table} 

Finite spectral triples (and consequently almost-commutative geometries) can be classified using Krajewski diagrams \cite{KR97}, which consist of a grid (labeled by the summands of the finite algebra) in which irreducible representations of $\A_F$ are placed as vertices. A component of the finite Dirac operator that maps between two particular representation spaces is then represented by an edge between the corresponding vertices. The existence of $J_F$ then implies that such a diagram is symmetric around its diagonal. The value for the finite grading $\gamma_F$ is represented by a $\pm$-sign inside the vertices.\footnote{In our nomenclature, an element with eigenvalue $
+ 1$ or $-1$ of a grading $\gamma_M$ / $\gamma_F$ is called \emph{left-handed} resp.~\emph{right-handed}, with corresponding subscripts $L,R$.} See the aforementioned references for details on Krajewski diagrams.\\

Dirac operators are seen to exhibit quite naturally what are called \emph{inner fluctuations} \cite{C96}, \cite[\S \MakeUppercase{\romannumeral 11}]{C00}; additional contributions that are of the form
\bas
	D \to D_A := D + A \pm JAJ^*,
\eas	
with $A$ self-adjoint and where the last term only arises when $D$ is part of a real spectral triple. For the canonical spectral triple these inner fluctuations are denoted by $D_\mu = \nabla^S_\mu + \mathbb{A}_\mu$, with $\mathbb{A}_\mu = -i\ad(g A_\mu)$ and $A_\mu$ a self-adjoint \emph{gauge field}. For a finite Dirac operator the inner fluctuations are generically denoted by $\Phi$ and are seen to give rise to scalar fields. \\

To each almost-commutative geometry we can associate a natural, gauge invariant action \cite{CC97}:
\begin{align}
	\frac{1}{2}\langle J\psi, D_A \psi \rangle + \tr f(D_A/\Lambda),\qquad \psi \in \H^+,\label{eq:totalaction}
\end{align}
where $f$ must be a positive, even function, $\Lambda$ is an (a priori unknown) mass scale and with $\langle ., .\rangle$ we denote the inner product on $\H = L^2(M, S\otimes \H_F)$, but restricted to spinors of $\gamma_M\otimes \gamma_F$--eigenvalue $+1$. This restriction is needed to avoid overcounting the fermionic degrees of freedom \cite{LMMS97, CCM07}, but requires $\gamma_M \otimes \gamma_F$ to anticommute with $J_M \otimes J_F$, i.e.~we require the KO-dimension of the full spectral triple to equal $2$ or $6$. We will restrict ourselves to four-dimensional manifolds and, following the success of the Standard Model from noncommutative geometry, demand the finite spectral triple to have KO-dimension 6. The second term of \eqref{eq:totalaction}, called the \emph{spectral action}, is in the context of almost-commutative geometries typically handled by performing a heat kernel expansion \cite{Gil84} in $\Lambda$. For almost-commutative geometries on compact, flat, four-dimensional manifolds without boundary ---the objects we are studying here--- the first terms of this expansion read \cite{CC97, BS13I}:
\begin{align}
\tr f\bigg(\frac{D_A}{\Lambda}\bigg) &\sim \int_M \bigg[\frac{f(0)}{8\pi^2}\Big( - \frac{1}{3}\tr_F\mathbb{F}_{\mu\nu}\mathbb{F}^{\mu\nu} + \tr_F \Phi^4 + \tr_F [D_\mu, \Phi]^2\Big)\nonumber \\
	&\qquad + \frac{1}{2\pi^2}\Lambda^4 f_4\tr_F \id - \frac{1}{2\pi^2}\Lambda^2 f_2\tr_F\Phi^2\bigg]  + \mathcal{O}(\Lambda^{-2}), \label{eq:spectral_action_acg_flat}
\end{align}
where $f_{n}$ is the $n$-th moment of the function $f$, with $\tr_F$ we mean the trace over the finite Hilbert space and $\mathbb{F}_{\mu\nu}$ denotes the (anti-Hermitian) field strength (or curvature) that corresponds to $\mathbb{A}_\mu$. Thus, physically, the Hilbert space $\H$ contains all fermionic data, the gauge bosons are generated by the canonical Dirac operator \dirac and the scalar fields are generated by $D_F$. \\

Krajewski diagrams can also be useful in determining what the action corresponding to a particular almost-commutative geometry is; contributions from the trace of the $n$th power of $D_F$ consist of all paths in the diagram consisting of $n$ steps and ending in the same point as where they started. Much like in the case of Feynman diagrams, the total contribution that corresponds to a particular path is the product of the factors that are associated to each of the edges it consists of.\\

\begin{figure}[h!]
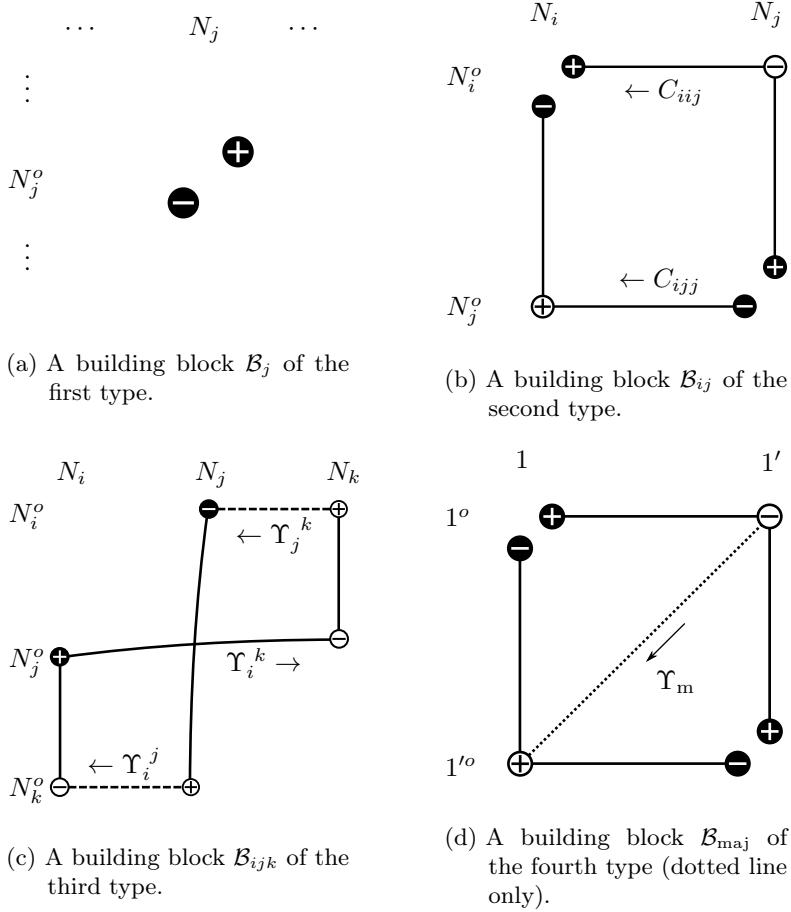

	\centering
	\begin{subfigure}{.3\textwidth}
		\centering
		\def\svgwidth{\textwidth}
		\subimport{\the\svgpath}{bb1.pdf_tex}
		\caption{A building block \B{j} of the first type.}
		\label{fig:bb1}
	\end{subfigure}
	\hspace{30pt}
\begin{subfigure}{.3\textwidth}
		\centering
		\def\svgwidth{\textwidth}
		\subimport{\the\svgpath}{bb2_Rplus.pdf_tex}
		\caption{A building block \B{ij} of the second type.}
		\label{fig:bb2}
	\end{subfigure}
	\\[10pt]
	\begin{subfigure}{.3\textwidth}
		\centering
		\def\svgwidth{\textwidth}
		\subimport{\the\svgpath}{bb3.pdf_tex}
		\caption{A building block \B{ijk} of the third type.}
		\label{fig:bb3}
	\end{subfigure}
	\hspace{30pt}
\begin{subfigure}{.3\textwidth}
		\centering
		\def\svgwidth{\textwidth}
		\subimport{\the\svgpath}{bb4.pdf_tex}
		\caption{A building block \BBBB{11'} of the fourth type (dotted line only).}
		\label{fig:bb4}
	\end{subfigure}

\captionsetup{width=.9\textwidth}
\caption{The Krajewski diagrams of four of the five different building blocks of almost-commutative geometries that potentially yield supersymmetric actions. The first corresponds to a gaugino--gauge boson pair, the second to a fermion--sfermion pair, the third to a superpotential interaction and the fourth to a Majorana mass for a gauge singlet. The fifth building block (not shown here) corresponds to a mass-like term for two fermions in the same representation of the gauge group. Note that the third block only contains the edges, not the vertices. The fourth block only contains the dotted edge. The sign inside the vertices represents their chirality. The white vertices correspond to fermions that have $R$-parity equal to $1$, i.e.~that are SM-like (and can consequently come in multiplicities higher than one). The black vertices have $R$-parity $-1$ and correspond to superpartners of bosons such as those of the SM.}
\label{fig:bbs}
\end{figure}

Thus, \emph{given an almost-commutative geometry}, the corresponding action is fixed. When we are talking about supersymmetric almost-commutative geometries, we thus mean those whose action is \emph{supersymmetric}, i.e.
\begin{align}\label{eq:susy_action_no_aux}
	\delta S[\sfer, \psi, \gau{}, A] &:=  \frac{\mathrm{d}}{\mathrm{d} t} S[\sfer + t\delta\sfer, \psi + t\delta \psi, \gau{} + t\delta\gau{}, A + t\delta A]\Big|_{t = 0} = 0.
\end{align} 
Here with $\sfer$, $\psi$, $\gau{}$ and $A$ we generically denote the respective \emph{sfermions} and \emph{Higgs scalars}, \emph{fermions} and \emph{higgsinos}, \emph{gauginos} and \emph{gauge bosons} of the theory and $\delta \zeta$ ($\zeta = \sfer, \psi, \gau{}, A$) are the \emph{supersymmetry transformations} that feature the superpartner of the respective field.\footnote{Equation \ref{eq:susy_action_no_aux} does not mention the auxiliary fields that are needed to ensure supersymmetry both on shell and off shell. The spectral action gives an on shell action and needs to be written off shell, introducing auxiliary fields $D$ and $F$. These then both appear in the action and in the supersymmetry transformations \cite{BS13I}.} The central result of \cite{BS13I} was that each non-commutative geometry that is fully decomposable in the five \emph{building blocks} \B{i}, \Bc{ij}{\pm}, \B{ijk}, \B{\mathrm{maj}} and \B{\mathrm{mass}} (the first four of which are depicted in Figure \ref{fig:bbs}) are eligible to have a supersymmetric action and do so when they satisfy certain additional demands (see \cite[\S 3]{BS13I}). Here, the building block \B{i} describes a gaugino--gauge boson pair in the adjoint representation of $SU(N_i)$ and corresponds to a vector multiplet in the parlance of superfields. The building block \Bc{ij}{\pm} of the second type (which requires building blocks \B{i}, \B{j} of the first type) describes a fermion--sfermion pair in the representation \rep{i}{j} with the fermion $\fer{ij}$ being left-handed ($+$) or right-handed ($-$), respectively. This corresponds to a chiral multiplet. The third building block \B{ijk} (requiring building blocks \Bc{ij}{\pm}, \Bc{ik}{\mp} and \Bc{jk}{\pm} of the second type) describes extra fermionic and bosonic interactions and corresponds to a term in a superpotential consisting of the product of three different chiral multiplets. The building block $\B{\mathrm{maj}}$ (requiring a singlet \B{11'}) corresponds to a Majorana mass for a gauge singlet. Finally the fifth building block $\B{\mathrm{mass}}$ (not depicted in Figure \ref{fig:bbs}, requiring two building blocks \B{ij} of opposite chirality) describes a mass-like term between two different fermions in the same representation.

\section{Soft supersymmetry breaking}\label{sec:Lsoft}

Consider a simple gauge group $G$, a set of scalar fields $\{\sfer_\alpha, \alpha = 1, \ldots, N\}$, all in a representation of $G$, and gauginos $\gau{} = \gau{a} T^a$, with $T^a$ the generators of $G$. Then the most general renormalizable Lagrangian that breaks supersymmetry softly is given \cite{GG81} by
\begin{align}
	\mathcal{L}_{\mathrm{soft}} &= - \sfer_\alpha^*(m^2)_{\alpha\beta} \sfer_\beta + \bigg(\frac{1}{3!} A_{\alpha\beta\gamma} \sfer_\alpha\sfer_\beta\sfer_\gamma - \frac{1}{2}B_{\alpha\beta}\sfer_\alpha\sfer_\beta + C_\alpha\sfer_\alpha + h.c.\bigg)\nonumber \\
	&\qquad- \frac{1}{2}(M\gau{a} \gau{a} + h.c.)\label{eq:LsoftM},
\end{align}
where the combinations of fields should be such that each term is gauge invariant. This expression contains the following terms:
\begin{itemize}
	\item mass terms for the scalar bosons $\sfer_\alpha$. For the action to be real, the matrix $m^2$ should be self-adjoint;
	\item trilinear couplings, proportional to a symmetric tensor $A_{\alpha\beta\gamma}$ of mass dimension $1$;
	\item bilinear scalar interactions via a matrix $B_{\alpha\beta}$ of mass dimension two;
	\item for gauge singlets there can be linear couplings, with $C_\alpha \in \mathbb{C}$ having mass dimension three;
	\item gaugino mass terms, with $M \in \com$.
\end{itemize}

It is important to note that the Lagrangian \eqref{eq:LsoftM} corresponds to a theory that is defined on a Minkowskian background. Performing a Wick transformation $t \to i\tau$ for the time variable to translate it to a theory on a Euclidean background, changes all the signs in \eqref{eq:LsoftM}:
\begin{align}
	\mathcal{L}_{\mathrm{soft}}^{\mathrm{E}} &=  \sfer_\alpha^*(m^2)_{\alpha\beta} \sfer_\beta - \bigg(\frac{1}{3!} A_{\alpha\beta\gamma} \sfer_\alpha\sfer_\beta\sfer_\gamma - \frac{1}{2}B_{\alpha\beta}\sfer_\alpha\sfer_\beta + C_\alpha\sfer_\alpha + h.c.\bigg)\nonumber \\
	&\qquad + \frac{1}{2}(M\gau{a} \gau{a} + h.c.)\label{eq:LsoftE}.
\end{align}
This expression can easily be extended to the case of a direct product of simple groups, but its main purpose is to give an idea of what soft supersymmetry breaking terms typically look like.   

\section{Soft supersymmetry breaking terms from the spectral action}\label{sec:terms}

As was mentioned in Section \ref{sec:recap}, we have to settle with the terms in the action that the spectral action principle provides us. The question at hand is thus whether noncommutative geometry can give us terms needed to break the supersymmetry. In \cite{BS13I} we have disregarded the second to last term ($\propto \Lambda^2$) in the expansion \eqref{eq:spectral_action_acg_flat} of the spectral action. Here we \emph{will} take this term into account.\\

In the following sections we will check for each of the terms in \eqref{eq:LsoftE} if it can also occur in the spectral action \eqref{eq:totalaction} (with \eqref{eq:spectral_action_acg_flat} for the expansion of its second term) in the context of the building blocks. We will denote scalar fields generically by $\sfer_{ij} \in C^{\infty}(M, \rep{i}{j})$, fermions by $\fer{ij} \in L^2(M, S\otimes \rep{i}{j})$ and gauginos by $\gau{i} \in L^2(M, S \otimes M_{N_i}(\com))$, with $M_{N_i}(\com) \to su(N_i)$ after reducing the gaugino degrees of freedom, \cite[\S 2.1.1]{BS13I}.

\subsection{Scalar masses (e.g.~Higgs masses)}\label{sec:breaking_scalar_mass}

\begin{figure}[ht]
	\begin{center}
		\def\svgwidth{.4\textwidth}
		\subimport{\the\svgpath}{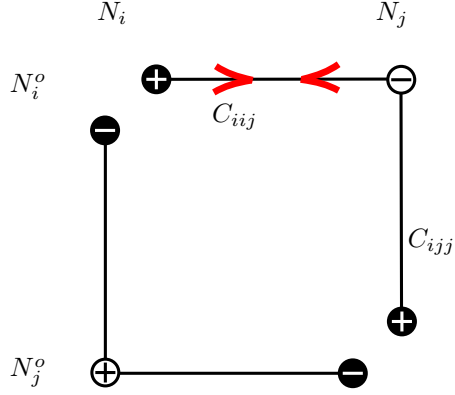}	
	\captionsetup{width=.6\textwidth}
\caption{A building block of the second type that defines a fermion--sfermion pair $(\fer{ij}, \sfer_{ij})$. Contributions to the mass term of the sfermion correspond to paths going back and forth on an edge, as is depicted on the top edge.}
	\label{fig:bb2-mass}
\end{center}
\end{figure}

Terms that describe the masses of the scalar particles such as the first term of \eqref{eq:LsoftE} are known \cite[\S 5.4]{KR97} to originate from the square of the finite Dirac operator (c.f.~\eqref{eq:spectral_action_acg_flat}). In terms of Krajewski diagrams these contributions are given by paths such as depicted in Figure \ref{fig:bb2-mass}.

Then the contribution to the action from a building block of the second type is:
\begin{align}
	- \frac{1}{2\pi^2}\Lambda^2 f_2\tr_F \Phi^2 =  - \frac{1}{2\pi^2} \Lambda^2f_2\big(4N_i|C_{iij}\sfer_{ij}|^2 + 4N_j|C_{ijj}\sfer_{ij}|^2\big)\label{eq:scalar-mass-term}
\end{align}
where $N_{i,j}$ are the dimensions of the representations $\mathbf{N_{i,j}}$ and $\sfer_{ij}$ is the field that is generated by the components of $D_F$ parametrized by $C_{iij}$ and $C_{ijj}$. Their expression depends on which building blocks are present in the spectral triple. 

In the case that there is a building block \B{ijk} of the third type present (parametrized by ---say--- $\yuk{i}{j}$, $\yuk{i}{k}$ and $\yuk{j}{k}$ acting on family-space),  we can both get the correct fermion--sfermion--gaugino interaction and a normalized kinetic term for the sfermion $\sfer_{ij}$ by on the one hand setting 
\ba
	C_{iij} = \sgnc_{i,j} \sqrt{\frac{r_i}{\w{ij}}}(N_k\yuks{i}{j}\yuk{i}{j})^{1/2}, \quad C_{ijj} = s_{ij} \sqrt{\frac{r_j}{r_i}}C_{iij},\quad s_{ij} = \sgnc_{i,j}\sgnc_{j,i} 
\label{eq:bb3-C}
\ea
where $\sgnc_{i,j}, \sgnc_{j,i}, s_{ij} \in \{\pm 1\}$, $r_i := q_in_i$ with $q_i:= f(0)g_i^2/\pi^2$, $n_i$ the normalization constant for the generators $T_i^a$ of $su(N_i)$ in the fundamental representation and $\w{ij} := 1 - r_iN_i - r_jN_j$. On the other hand we scale the sfermion according to
\ba
	\sfer_{ij} \to \n_{ij}^{-1}\sfer_{ij},\quad \text{with }\quad \n_{ij}^{-1} = \sqrt{\frac{2\pi^2\w{ij}}{f(0)}}(N_k\yuks{i}{j}\yuk{i}{j})^{-1/2}.\label{eq:bb3-scale}
\ea
There is an extra contribution from $\tr_F \Phi^2$ to $|\sfer_{ij}|^2$ compared to that of the building block of the second type. This contribution corresponds to paths going back and forth over the rightmost and bottommost edges in Figure \ref{fig:bb3}. In the parametrizations \eqref{eq:bb3-C} and upon scaling according to \eqref{eq:bb3-scale} these together yield 
\ba
 - \frac{1}{2\pi^2} \Lambda^2f_2\Big(4N_i|C_{iij}\sfer_{ij}|^2 + 4N_j|C_{ijj}\sfer_{ij}|^2 + 4N_k|\yuk{i}{j}\sfer_{ij}|^2\Big) &\to - 4 \Lambda^2\frac{f_2}{f(0)}|\sfer_{ij}|^2 \label{eq:scalar-mass-term3},
\ea
and similar expressions for $|\sfer_{ik}|^2$ and $|\sfer_{jk}|^2$. Interestingly, the pre-factor for this contribution is universal, i.e.~it is completely independent from the representation \rep{i}{j} the scalar resides in.

Note that, for $\Lambda \in \mathbb{R}$ and $f(x)$ a positive function (as is required for the spectral action) in both cases the scalar mass contributions are of the wrong sign, i.e.~they have the same sign as a Higgs-type scalar potential would have. The result would be a theory whose gauge group is broken maximally. We will see that, perhaps counterintuitively, we can escape this by adding gaugino-masses.

	\subsection{Gaugino masses}\label{sec:breaking_gaugino}

Having a building block of the first type, that consists of two copies of $M_N(\com)$ for a particular value of $N$, allows us to define a finite Dirac operator whose two components map between these copies, since both are of opposite grading. On the basis $\H_F = M_N(\com)_L \oplus M_N(\com)_R$ this is written as
\begin{align*}
	D_F = \begin{pmatrix} 0 & G \\ G^* & 0 \end{pmatrix},\qquad G : M_{N}(\com)_R \to M_{N}(\com)_L,
\end{align*}
since it needs to be self-adjoint. This form for $D_F$ automatically satisfies the order one condition \eqref{eq:order_one} and the demand $JD = DJ$ (see \eqref{eq:JD-anti-comm}) translates into $G = JG^*J^*$. If we want this to be a genuine mass term it should not generate any scalar field via its inner fluctuations. For this $G$ must be a multiple of the identity and consequently we write $G = M \id_{N}$, $M \in \com$. This particular pre-factor is dictated by how the term appears in \eqref{eq:LsoftE}.

For the fermionic action we then have
\begin{align}
	\frac{1}{2}\langle J (\gau{L}, \gau{R}), \gamma^5 D_F(\gau{L}, \gau{R})\rangle =  
\frac{1}{2}M\langle J_M\gau{R}, \gamma^5 \gau{R}\rangle + \frac{1}{2}\overline{M}\langle J_M\gau{L}, \gamma^5 \gau{L}\rangle, \label{eq:gaugino-mass}
\end{align}
where $(\gau{L}, \gau{R}) \in \H^+ = L^2(S_+ \otimes M_{N}(\com)_L) \oplus L^2(S_- \otimes M_{N}(\com)_R)$, with $S_{\pm}$ the space of left- resp.~right-handed spinors. This indeed describes a gaugino mass term for a theory on a Euclidean background (cf.~\cite{CCM07}, equation 4.52).

\begin{figure}[ht]
\begin{center}
		\def\svgwidth{.4\textwidth}
		\subimport{\the\svgpath}{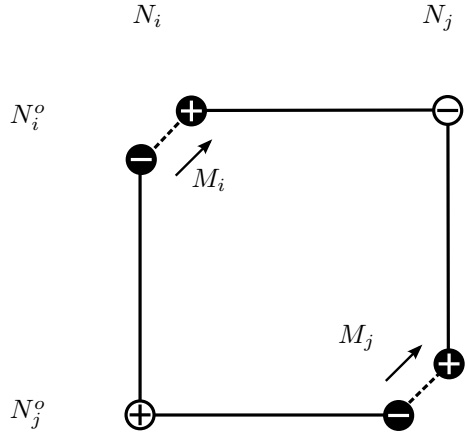}
\captionsetup{width=.9\textwidth}
\caption{A building block of the second type that defines a fermion--sfermion pair $(\fer{ij}, \sfer_{ij})$, dressed with mass terms for the corresponding gauginos (dashed edges, labeled by $M_{i,j}$).}
\label{fig:gaugMass}
\end{center}
\end{figure}

A gaugino mass term in combination with building blocks of the second type (for which two gaugino pairs are required), gives extra contributions to the spectral action. From the set up as is depicted in Figure \ref{fig:gaugMass}, one can see that $\tr D_F^4$ receives extra contributions coming from paths that traverse two edges representing a gaugino mass and two representing the scalar $\sfer_{ij}$. In detail, the extra contributions are given by:
\begin{align}
	  \frac{f(0)}{8\pi^2} \tr_F \Phi^4 &= \frac{f(0)}{\pi^2}\big(N_i|M_i|^2|C_{iij}\sfer_{ij}|^2 + N_j|M_j|^2|C_{ijj}\sfer_{ij}|^2\big)\nonumber\\
	 						&\to 2\Big(r_iN_i|M_i|^2 + r_jN_j|M_j|^2\Big)|\sfer_{ij}|^2. 
\label{eq:scalar-mass-gaugino}
\end{align}
upon scaling the fields.
 
This means that there is an extra contribution to the scalar mass terms, that is of opposite sign (i.e.~positive) as compared to the one from the previous section. When
\bas
	2r_iN_i|M_i|^2 + 2r_jN_j|M_j|^2 > 4\frac{f_2}{f(0)}\Lambda^2,
\eas 
then the mass terms of the sfermions have the correct sign, averting the problem of a maximally broken gauge group that was mentioned in the previous section. Comparing this with the expression for the Higgs mass(es) raises interesting questions about the physical interpretation of this result. In particular, if we would require the mass terms of the sfermions and Higgs boson(s) to have the correct sign already at the scale $\Lambda$ on which we perform the expansion of the spectral action, this seems to suggest that at least some gaugino masses must be very large.

Note that a gauge singlet $\fer{\mathrm{sin}} \in L^2(M, S\otimes \repl{1}{1'})$ (such as the right-handed neutrino) can be dressed with a Majorana mass matrix $\maj$ in family space (see \cite[\S 2.6]{CCM07} and Figure \ref{fig:singlet}). This yields extra supersymmetry breaking contributions: 
\begin{align}
	&\frac{f(0)}{8\pi^2}\tr \Big[4(C_{111'}\sfer_{\mathrm{sin}})\overline{\maj}(C_{111'}\sfer_{\mathrm{sin}})\overline{M} + 4(C_{11'1'}\sfer_{\mathrm{sin}})\overline{\maj}(C_{11'1'}\sfer_{\mathrm{sin}})\overline{M'}\Big] + h.c.  \nn\\
	&\qquad \to r_1(\overline{M} + \overline{M'})\tr \overline{\maj}\sfer_{\mathrm{sin}}^2 + h.c. \label{eq:singletterm}
\end{align}
where $M$ and $M'$ denote the gaugino masses of the two one-dimensional building blocks \B{1}, \B{1'} of the first type respectively and the trace is over family space. This expression is independent of whether there are building bocks of the third type present.

Note furthermore that the gaugino masses do not give rise to mass terms for the gauge bosons. In the spectral action such terms could come from an expression featuring both $D_A = i \gamma^\mu D_\mu $ and $D_F$ twice. We do have such a term in \eqref{eq:spectral_action_acg_flat} but since it appears with a commutator between the two and since we demanded the gaugino masses to be a multiple of the identity in $M_N(\com)$, such terms vanish automatically. (In contrast, the Higgs boson does generate mass terms for the $W^{\pm}$- and $Z$-bosons, partly since the Higgs is not in the adjoint representation.)

\subsection{Linear couplings} \label{sec:breaking_lin}

The fourth term of \eqref{eq:LsoftE} can only occur for a gauge singlet, i.e.~the representation \repl{1}{1} (or, quite similarly, the representation \repl{\overline{1}}{\overline{1}}). The only situation in which such a term can arise is with a building block of the second type --- defining a fermion--sfermion pair $(\fer{\mathrm{sin}}, \sfer_{\mathrm{sin}})$ and their antiparticles (see Figure \ref{fig:singlet}). Moreover in this case a Majorana mass $\maj$ is possible, that does not generate a new field.

\begin{figure}[ht]
\begin{center}
		\def\svgwidth{.4\textwidth}
		\captionsetup{width=.9\textwidth}
		\subimport{\the\svgpath}{bb2-maj.pdf_tex}	
\caption{A building block of the second type that defines a gauge singlet fermion--sfermion pair $(\fer{\mathrm{sin}}, \sfer_{\mathrm{sin}})$. Moreover, a Majorana mass term $\maj$ is possible.}
\label{fig:singlet}
\end{center}
\end{figure}

Any such term in the spectral action must originate from a path in this Krajewski diagram consisting of either two or four steps (corresponding to the second and fourth power of the Dirac operator), ending at the same vertex at which it started (if it is to contribute to the trace) and traversing an edge labeled by $\sfer_{\mathrm{sin}}$ only once. From the diagram one readily checks that such a contribution cannot exist.

	\subsection{Bilinear couplings}\label{sec:breaking_bilin}

\begin{figure}[ht]
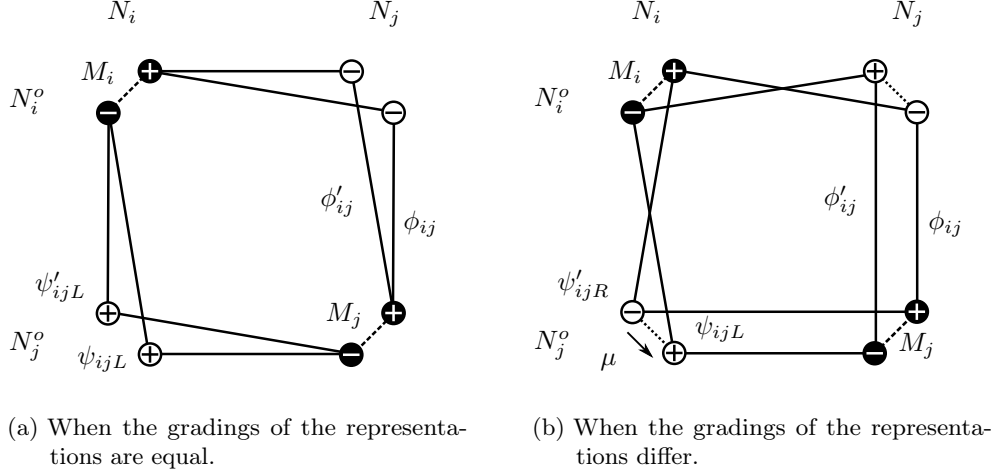

	\centering
	\begin{subfigure}{.40\textwidth}
		\centering
		\def\svgwidth{\textwidth}
		\subimport{\the\svgpath}{bilin1.pdf_tex}
		\caption{When the gradings of the representations are equal.}
		\label{fig:bilin1}
	\end{subfigure}
	\hspace{20pt}
	\begin{subfigure}{.40\textwidth}
		\centering
		\def\svgwidth{\textwidth}
		\subimport{\the\svgpath}{bilin2.pdf_tex}
		\caption{When the gradings of the representations differ.}
		\label{fig:bilin2}
	\end{subfigure}
		\captionsetup{width=.9\textwidth}
		\caption{Two building blocks of the second type defining two fermion--sfermion pairs $(\fer{ij}, \sfer_{ij})$ and $(\fer{ij}', \sfer_{ij}')$ in the same representation.} 
		\label{fig:bilin}	
\end{figure}

If a bilinear coupling (such as the third term in \eqref{eq:LsoftE}) is to be a gauge singlet, the two fields $\sfer_{ij}$ and $\sfer_{ij}'$ appearing in the expression should have opposite finite representations, e.g.~$\sfer_{ij} \in C^\infty(M, \rep{i}{j})$, $\sfer_{ij}' \in C^\infty(M, \rep{j}{i})$. We will rename $\sfer_{ij}' \to \asfer_{ij}'$ for consistency with \cite[\S 2.5.2]{BS13I}. The building blocks of the second type by which they are defined are depicted in Figure \ref{fig:bilin}.

The gradings of both representations are either the same (left image of Figure \ref{fig:bilin}), or they are of opposite eigenvalue (the right image). A contribution to the action that resembles the third term in \eqref{eq:LsoftE} needs to come from paths in the Krajewski diagram of Figure \ref{fig:bilin} consisting of either two or four steps, ending in the same point as where they started and traversing an edge labeled by $\sfer_{ij}$ and $\sfer_{ij}'$ only once.

One can easily check that in the left image of Figure \ref{fig:bilin} no such paths exist. In the second case (right image of Figure \ref{fig:bilin}), however, there arises the possibility of a component $\mu$ of the finite Dirac operator that maps between the vertices labeled by $\fer{ij}$ and $\fer{ij}'$ (and consequently also between $\afer{ij}$ and $\afer{ij}'$). This corresponds to a building block of the fifth type (Section \ref{sec:recap}). There is a contribution to the action (via $\tr D_F^4$) that comes from loops traversing both an edge representing a gaugino mass and one representing $\mu$. If the component $\mu$ is parameterized by a complex number, then the contribution is 
\begin{align}	
	& \frac{f(0)}{8\pi^2} \big(8 N_i\tr M_i\asfer_{ij}C_{iij}^*\mu C_{iij}'\sfer_{ij}' + 8N_j\tr M_j \asfer_{ij}C_{ijj}^* \mu C_{ijj}'\sfer_{ij}'\big) + h.c.\nn\\
	&\qquad \to 2 \big(r_iN_iM_i + r_jN_jM_j\big)\mu \tr\asfer_{ij}\sfer_{ij}' + h.c.\label{eq:bilinear},
\end{align}
where the traces are over $\mathbf{N}_j^{\oplus M}$, with $M$ the number of copies of \rep{i}{j}. This indeed yields a bilinear term such as the third one of \eqref{eq:LsoftE}. 

\subsection{Trilinear couplings}\label{sec:breaking_trilin}

	Trilinear terms such as the second term of \eqref{eq:LsoftE} might appear in the spectral action. For that we need three fields $\sfer_{ij} \in C^{\infty}(M, \rep{i}{j})$, $\sfer_{jk} \in C^{\infty}(M, \rep{j}{k})$ and $\sfer_{ik} \in C^{\infty}(M, \rep{i}{k})$, generated by the finite Dirac operator. Such a term can only arise from the fourth power of the finite Dirac operator\footnote{Here we assume that each component of the finite Dirac operator generates only a single field, instead of ---say--- two composite ones.} which is visualized by paths in the Krajewski diagram consisting of four steps, three of which correspond to a component that generates a scalar field, the other one must be a term that does not generate inner fluctuations, e.g.~a mass term. Non-gaugino fermion mass terms were already covered in \cite{BS13I} and were seen to generate potentially supersymmetric trilinear interactions, so the mass term must be a gaugino mass.

\begin{figure}[ht]
\begin{center}
		\def\svgwidth{.5\textwidth}
		\subimport{\the\svgpath}{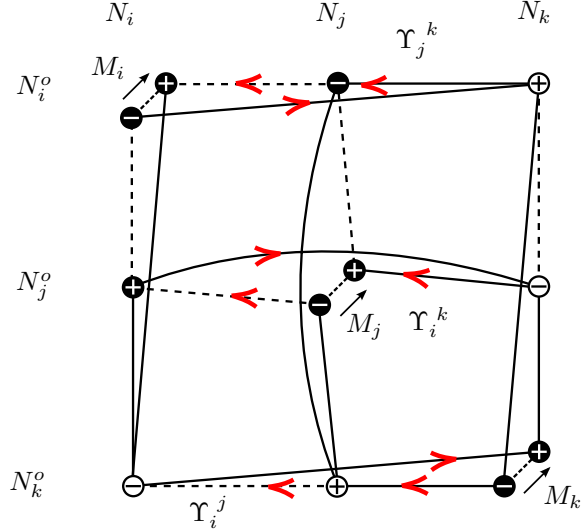}	
	\captionsetup{width=.9\textwidth}
\caption{A situation in which there are three building blocks $\mathcal{B}_{i,j,k}$ of the first type (black vertices), three building blocks $\mathcal{B}_{ij,jk,ik}$ of the second type and a building block $\mathcal{B}_{ijk}$ of the third type. Adding gaugino masses (dashed edges) gives rise to trilinear interactions, corresponding to the paths in the diagram marked by arrows.}
\label{fig:trilinear}
\end{center}
\end{figure}

If the component of the finite Dirac operator that does not generate a field is a gaugino mass term (mapping between ---say--- $M_{N_i}(\com)_R$ and $M_{N_i}(\com)_L$), then two of the three components that do generate a field must come from building blocks of the second type, since they are the only ones connecting to the adjoint representations. If we denote the non-adjoint representations from these building blocks by \rep{i}{j} and \rep{i}{k} then we can only get a contribution to $\tr D_F^4$ if there is a component of $D_F$ connecting these two representations. If $\mathbf{N_j} = \mathbf{N_k}$, such a component could yield a mass term for the fermion in the representation \rep{i}{j}, and we revert to the previous section. If $\mathbf{N_j} \ne \mathbf{N_k}$ then the remaining component of $D_F$ must be part of a building block of the third type, namely $\mathcal{B}_{ijk}$. This situation is depicted in Figure~\ref{fig:trilinear}. It gives rise to three different trilinear interactions corresponding to the paths labeled by arrows in the figure. Each of these three paths actually represents four contributions: one can traverse each path in the opposite direction, and for each path one can reflect it around the diagonal, giving another path with the same contribution to the action. 

Calculating the spectral action we get for each building block $\mathcal{B}_{ijk}$ of the third type the contributions
\begin{align}
	&\frac{f(0)}{\pi^2}\Big(N_i\overline{M_i}\tr \yuk{j}{k}\sfer_{jk}\asfer_{ik}C_{iik}^*C_{iij}\sfer_{ij} + N_j\overline{M_j}\tr C_{jjk}\sfer_{jk}\asfer_{ik}\yuk{i}{k}C_{ijj}\sfer_{ij} \nn\\
	&\qquad\qquad + N_k\overline{M_k}\tr C_{jkk}\sfer_{jk}\asfer_{ik}C_{ikk}^*\yuk{i}{j}\sfer_{ij}\Big) + h.c.\label{eq:trilinear_prior}
\end{align}
where all traces are over $\mathbf{N}_j^{\oplus M}$. A careful analysis of the demand for supersymmetry in this context (see \cite[\S 2.3]{BS13I}) requires the parameters $\yuk{i}{j}$, $\yuk{i}{k}$ and $\yuk{j}{k}$ to be related via 
\ba
C_{ikk}^*\yuk{j}{k} &= - \yuk{i}{k}C_{jkk}, & \yuk{i}{k} C_{iij} &= - C_{iik}^* \yuk{i}{j}, & \yuk{i}{j} C_{jjk} &= - \yuk{j}{k}C_{ijj}
\label{eq:improvedUpsilons1},
\ea
where $C_{iij}$ and $C_{ijj}$ act trivially on family space if $\sfer_{ij}$ is assumed to have $R =1$. From this relation we can deduce that $s_{ij}s_{ik}s_{jk} = -1$ for the product of the three signs defined in \eqref{eq:bb3-C}. If we replace $C_{iik} \to C_{ikk}$, $C_{iij} \to C_{ijj}$, $C_{jjk} \to C_{jkk}$ and $C_{ijj} \to C_{iij}$ in the first two terms of \eqref{eq:trilinear_prior} using \eqref{eq:bb3-C}, employ \eqref{eq:improvedUpsilons1}, then \eqref{eq:trilinear_prior} can be written as
\bas
&\frac{f(0)}{\pi^2}\Big(N_i\overline{M_i}\frac{r_i}{r_k} + N_j\overline{M_j}\frac{r_j}{r_k}	+ N_k\overline{M_k}\Big)\tr C_{jkk}\sfer_{jk}\asfer_{ik}C_{ikk}^*\yuk{i}{j}\sfer_{ij} + h.c.
\eas
We then scale the sfermions according to \eqref{eq:bb3-scale}, again using \eqref{eq:bb3-C} for $C_{jkk}$ and $C_{ikk}^*$ to obtain 
\begin{align}
	& 2\kappa_k g_l\sqrt{2\frac{\w{ij}}{q_l}}\Big(r_iN_i\overline{M_i} + r_jN_j\overline{M_j} + r_kN_k\overline{M_k}\Big)\tr \yukw{}{}\sfer_{ij}\sfer_{jk}\asfer_{ik} + h.c.\label{eq:trilinear},
\end{align}
where we have written
\bas
	\yukw{}{} := \yuk{i}{j}(N_k\tr\yuks{i}{j}\yuk{i}{j})^{-1/2}
\eas
for the scaled version of the parameter $\yuk{i}{j}$, $\kappa_k := \sgnc_{k,j}\sgnc_{k,i}$ and the index $l$ can take any of the values that appear in the theory.



\section{Conclusion}

We have now considered all terms featuring in \eqref{eq:LsoftE}. At the same time the reader can convince himself that this exhausts all possible terms that appear via $\tr D_F^4$ and feature a gaugino mass. As for the fermionic action, a component of $D_F$ mapping between two adjoint representations can give gaugino mass terms \eqref{eq:gaugino-mass}. As for the bosonic action, any path of length two contributing to the trace and featuring a gaugino mass, cannot feature other fields. In contrast, a path of length four in a Krajewski diagram involving a gaugino mass can feature:
\begin{itemize}
	\item only that mass, as a constant term (see the comment at the end of this section);
	\item two times the scalar from a building block of the second type, when going in one direction \eqref{eq:scalar-mass-gaugino};
	\item two times the scalar from a building block of the second type, when going in two directions and when a Majorana mass is present (only possible for singlet representations, \eqref{eq:singletterm});
	\item two scalars from two different building blocks of the second type having opposite grading in combination with a building block of the fifth type \eqref{eq:bilinear}.
 	\item three scalars, partly originating from a building block of the second type and partly from one of the third type \eqref{eq:trilinear}.
\end{itemize}
Furthermore, via $\tr D_F^2$ there are contributions to the scalar masses from building blocks of the second and third type \eqref{eq:scalar-mass-term}. We can combine the main results of the previous sections into the following theorem.
\begin{theorem}
All possible terms that break supersymmetry softly and that can originate from the spectral action \eqref{eq:spectral_action_acg_flat} of an almost-commutative geometry consisting of building blocks are mass terms for scalar fields and gauginos and trilinear and bilinear couplings. More precisely, the most general Lagrangian that softly breaks supersymmetry and results from almost-commutative geometries is of the form
\begin{align}
	\mathcal{L}_{\mathrm{soft}}^{\mathrm{NCG}} &= \mathcal{L}^{(1)} + \mathcal{L}^{(2)} + \mathcal{L}^{(3)}+ \mathcal{L}^{(4)} + \mathcal{L}^{(5)}\label{eq:main-result},
\end{align}
where
\begin{subequations}
\begin{align}
			\mathcal{L}^{(1)}&= \frac{1}{2}M_i \langle J_M\gau{iR}, \gamma^5\gau{iR} \rangle + \frac{1}{2}\overline{M_i} \langle J_M\gau{iL}, \gamma^5\gau{iL} \rangle \label{eq:gauginoterm}\\
	\intertext{for each building block \ensuremath{\mathcal{B}_i} of the first type,}
	\mathcal{L}^{(2)} &= 2\Big(r_iN_i|M_i|^2 + r_jN_j|M_j|^2 - 2\frac{f_2}{f(0)}\Lambda^2\Big)|\sfer_{ij}|^2,\label{eq:massterm2}
	\intertext{for each building block \ensuremath{\mathcal{B}_{ij}} of the second type for which there is at least one building block \B{ijk} of the third type present (knowing that a single \B{ij} cannot be supersymmetric by itself, \cite[\S 2.2]{BS13I}),}
	\mathcal{L}^{(3)} &= 
	  2\kappa_k g_l\sqrt{2\frac{\w{ij}}{q_l}}\Big(r_iN_i\overline{M_i} + r_jN_j\overline{M_j} + r_kN_k\overline{M_k}\Big)\tr \yukw{}{}\sfer_{ij}\sfer_{jk}\asfer_{ik} + h.c.
\label{eq:trilinterm},
	\intertext{for each building block \ensuremath{\mathcal{B}_{ijk}} of the third type,}
\mathcal{L}^{(4)} &= r_1(\overline{M} + \overline{M'})\tr \overline{\maj}\sfer_{\mathrm{sin}}^2+ h.c. \label{eq:majterm}
	\intertext{for each building block \ensuremath{\mathcal{B}_{\mathrm{maj}}} of the fourth type (with the trace over a possible family index), and}
	\mathcal{L}^{(5)} &= 2(r_iN_iM_{i} + r_jN_jM_{j})\mu \tr\asfer_{ij}\sfer_{ij}' + h.c.\label{eq:bilinterm}
\end{align}
\end{subequations}
for each building block $\mathcal{B}_{\mathrm{mass}}$ of the fifth type.
\end{theorem}
It should be remarked that the building blocks of the fourth and fifth type typically already provide soft breaking terms of their own (see \cite{BS13I}, Section 2.5).

Interestingly, all supersymmetry breaking interactions that occur are seen to be generated by the gaugino masses (except the ones coming from the trace of the square of the finite Dirac operator) and each of them can be associated to one of the five supersymmetric building blocks. Note that the gaugino masses give rise to extra contributions that are not listed in \eqref{eq:main-result}. For each gaugino mass $M_i$ there is an additional contribution 
\begin{align*}
\mathcal{L}_{M_i} &= \frac{f(0)}{4\pi^2}|M_i|^4 - \frac{f_2}{\pi^2}\Lambda^2|M_i|^2.
\end{align*}
Since such contributions do not contain fields, they are not breaking supersymmetry, but might nonetheless be interesting from a gravitational perspective.

\section*{Acknowledgements}
One of the authors would like to thank the Dutch Foundation for Fundamental Research on Matter (FOM) for funding this work.

\bibliographystyle{plain}

\begin{thebibliography}{10}

\bibitem{BS13I}
W.~Beenakker, T.~van~den Broek, and W.D. van Suijlekom.
\newblock Noncommutative geometry and supersymmetry. {P}art \rnum{1}:
  {S}upersymmetric almost-commutative geometries.
\newblock 2014.

\bibitem{CC97}
A.H. Chamseddine and A.~Connes.
\newblock The spectral action principle.
\newblock {\em Comm. Math. Phys.}, 186:731--750, 1997.

\bibitem{CCM07}
A.H. Chamseddine, A.~Connes, and M.~Marcolli.
\newblock Gravity and the standard model with neutrino mixing.
\newblock {\em Adv. Theor. Math. Phys.}, 11:991--1089, 2007.

\bibitem{C94}
A.~Connes.
\newblock {\em Noncommutative geometry}.
\newblock Academic Press, 1994.

\bibitem{C96}
A.~Connes.
\newblock Gravity coupled with matter and the foundation of noncommutative
  geometry.
\newblock {\em Commun. Math. Phys.}, 182:155--176, 1996.

\bibitem{C00}
A.~Connes.
\newblock Noncommutative geometry year 2000.
\newblock {\em math/0011193}, 2007.

\bibitem{Dabrowski2010}
L.~D\k{a}browski and G.~Dossena.
\newblock Product of real spectral triples.
\newblock {\em Int. J. Geom. Methods Mod. Phys.}, 8(8):1833--1848, 2010.

\bibitem{DGR04}
M.~Drees, R.~Godbole, and P.~Roy.
\newblock {\em Theory and phenomenology of Sparticles}.
\newblock World Scientific Publishing Co., 2004.

\bibitem{DS11}
K.~{\noopsort{dungen}}van~den Dungen and W.D. van Suijlekom.
\newblock Electrodynamics from noncommutative geometry.
\newblock {\em J. Noncommut. Geom.}, 7:433--456, 2013.

\bibitem{FI74}
P.~Fayet and J.~Iliopoulos.
\newblock Spontaneously broken supergauge symmetries and {G}oldstone spinors.
\newblock {\em Phys.~Lett.~B}, 51:461--464, 1974.

\bibitem{GD81}
H.~Georgi and S.~Dimopoulos.
\newblock Softly broken supersymmetry and ${SU(5)}$.
\newblock {\em Nucl.~Phys.~B}, 193:150 -- 162, 1981.

\bibitem{Gil84}
P.B. Gilkey.
\newblock {\em Invariance theory, the heat equation and the {A}tiyah-{S}inger
  index theorem}, volume~11 of {\em Mathematics Lecture Series}.
\newblock Publish or Perish, Wilmington, DE, 1984.

\bibitem{GG81}
L.~Girardello and M.T. Grisaru.
\newblock Soft breaking of supersymmetry.
\newblock {\em Nuclear Phys. B Proc. Suppl.}, 194:65 -- 76, 1981.

\bibitem{GVF00}
J.M. Gracia-Bond\'ia, J.C. V\'arilly, and H.~Figueroa.
\newblock {\em Elements of Noncommutative Geometry}.
\newblock Birkh\"auser {A}dvanced {T}exts, 2000.

\bibitem{GH86}
J.F. Gunion and H.E. HABER.
\newblock Higgs bosons in supersymmetric models (l).
\newblock {\em Nuclear Phys. B}, 272:1, 1986.

\bibitem{KR97}
T.~Krajewski.
\newblock Classification of finite spectral triples.
\newblock {\em J. Geom. Phys.}, 28:1--30, 1998.

\bibitem{LMMS97}
F.~Lizzi, G.~Mangano, G.~Miele, and G.~Sparano.
\newblock Fermion {H}ilbert space and fermion doubling in the noncommutative
  geometry approach to gauge theories.
\newblock {\em Phys. Rev. D}, 55:6357--6366, 1997.

\bibitem{R75}
L.~O'Raifeartaigh.
\newblock Spontaneous symmetry breaking for chiral scalar superfields.
\newblock {\em Nuclear Phys. B Proc. Suppl.}, 96:331--352, 1975.

\bibitem{S00}
T.~Sch\"ucker.
\newblock Spin group and almost commutative geometry.
\newblock {\em hep-th/0007047}, 2007.

\bibitem{WZ74-2}
J.~Wess and B.~Zumino.
\newblock Supergauge invariant extension of quantum electrodynamics.
\newblock {\em Nuclear Phys. B Proc. Suppl.}, 78:1--13, 1974.

\bibitem{WestBook}
P.~West.
\newblock {\em Introduction to Supersymmetry and Supergravity}.
\newblock World Scienctific Publishing, 2nd edition, 1990.

\end{thebibliography}
\providecommand{\noopsort}[1]{}

\end{document}